\documentclass{icrc29}
\usepackage{graphicx,amssymb,amsmath,times}
\begin{document}

\title[$\nu$-induced cascades from GRBs with AMANDA-II]{Neutrino-Induced
  Cascades From GRBs With AMANDA-II}
\author[B.~Hughey, I.~Taboada for the IceCube Collaboration] {B.~Hughey$^a$,
  I.~Taboada$^b$ for the IceCube Collaboration$^c$\\
        (a) Physics Dept. University of Wisconsin. Madison, WI 53706, USA \\
        (b) Physics Dept. University of California. Berkeley, CA 94720, USA\\
        (c) For a full author list see arXiv:astro-ph/0509330 \\
        }
\presenter{Presenter: B.~Hughey (brennan.hughey@icecube.wisc.edu), \
  usa-hughey-B-abs1-og25-oral}

\maketitle

\begin{abstract}
Using AMANDA-II we have performed a search for $\nu$-induced
cascades in coincidence with 73 bursts reported by BATSE in 2000.
Background is greatly suppressed by the BATSE temporal constraint. No
evidence of neutrinos was found. We set a limit on a WB-like spectrum, 
$A_{90}^{\rm all\;
  flavors}$~=~9.5$\times$10$^{-7}$~GeV~cm$^{-2}$~s$^{-1}$~sr$^{-1}$. 
The determination of systematic uncertainties is in progress, and the limit
will be somewhat weakened once these uncertainties are taken into account.
We are also conducting a rolling time-window search for $\nu$-induced
cascades consistent with a GRB signal in 2001. The data set is searched for a
statistically significant cluster of signal-like events within a 1~s or 100~s
time window.  The non-triggered search has the
potential to discover phenomena, including gamma-ray dark choked bursts, which
did not trigger gamma-ray detectors.
\end{abstract}

\section{Introduction}

Gamma Ray Bursts are among the most energetic processes in the universe. High
energy neutrinos ($\approx 10^{14}$~eV) are thought 
to be produced via the process $p+\gamma \rightarrow \Delta^+ \rightarrow \pi^+
[+n] \rightarrow \mu^+ + \nu_\mu \rightarrow e^+ + \bar{\nu}_e +
\nu_\mu$. Neutrino oscillations result in a flavor flux ratio, 
$\phi_{\nu_e+\bar{\nu}_e}$:$\phi_{\nu_\mu+\bar{\nu}_\mu}$:$\phi_{\nu_\tau+\bar{\nu}_\tau}$,
equal to 1:1:1 at Earth\footnote{But the ratio $\phi_\nu$:$\phi_{\bar{\nu}}$ is
  not 1:1.}.
AMANDA-II, a sub-detector of IceCube, was commissioned in 2000 with a total of
677 optical modules arranged on 19 strings, at depths between 1500~m and
2000~m below the surface of the ice at the South Pole. Each OM contains a
20~cm photo-multiplier tube in a pressure vessel. AMANDA-II uses polar ice as a
Cherenkov medium. Searches for $\nu$-induced muons with
AMANDA-II~\cite{ama:nature-pub,nu2004:ama} in coincidence with bursts 
reported by satellites have been done for 1997-2003~\cite{icrc03:grb}. These
searches take advantage of the spatial and temporal  localization of the
bursts to reduce background, but are restricted to bursts with positive
declination because AMANDA-II, located at the South Pole, relies on the use of
the Earth to filter out all non-neutrino particles from the northern
hemisphere.
The cascade channel is complementary to the muon channel. AMANDA-II is uniformly
sensitive to cascades from all directions, so objects at any declination can
be studied. Further, GRBs without directional information can be used as no
correlation to the cascade direction is required. Even though the detector's
effective volume is smaller for cascades than for muons, more bursts can be
studied with the cascade channel.
Isolated cascades are produced by several interactions: $\nu_e N$
charged current, $\nu_x N$ neutral current, $\bar{\nu}_e e^-$ at 6.3 PeV
(Glashow resonance) and $\nu_\tau N$ charged current in the case when the
$\tau$ travels a short distance before decaying and the decay cascade overlaps
the $\nu_\tau N$ hadronic cascade. A 100~TeV $\tau$  will travel $O(5~m)$
before decaying. As a comparison, a 100 TeV electromagnetic cascade is
$\approx$8.5~m long in ice.
 
We present two analyses that search for $\nu$-induced
cascades in coincidence with GRBs. For the first analysis, hereafter referred
to as the \textit{Rolling} analysis, we do not use any correlation with bursts reported by satellites. Instead
two time windows, 1~s and 100~s, are rolled along the data taken by
AMANDA-II in the year 2001, to search for statistical excess. This technique has
the advantage of being sensitive to bursts that were not reported by 
satellites. The second analysis, hereafter referred to as the
\textit{Temporal} analysis, uses the temporal, but not the spatial,
correlation with bursts reported by BATSE~\cite{batse} in the year 2000. Using
this correlation reduces the background significantly.

\section{Simulation and Reconstruction}

For both analyses neutrino-induced cascades for all three neutrino
flavors were simulated with \texttt{ANIS}~\cite{anis} from 100~GeV to 100~PeV
following an $E^{-1}$ spectrum. This simulation was then re-weighted to 
follow the flux predicted by the Waxman-Bahcall model~\cite{wb}. This spectrum
is derived from average burst 
characteristics, and thus it is adequate to describe a large number of bursts
simultaneously. Individual burst spectra may deviate significantly from the WB
spectrum. We use a break energy, $E_B$=100~TeV and a synchrotron energy,
$E_s$=10~PeV. For the Rolling analysis signal
simulation was verified with \texttt{TEA}~\cite{tea} which produces a
Waxman-Bahcall type broken power law spectrum directly. The outputs of
\texttt{ANIS} and \texttt{TEA} were found to be consistent. In both
the Rolling and Temporal analyses, background muon events were simulated
with \texttt{CORSIKA}~\cite{corsika}. Muons were propagated through ice using
\texttt{MMC}~\cite{MMC} and detector response was simulated with
\texttt{AMASIM}~\cite{amasim} for both signal and background simulation.   

For both analyses, data were reconstructed with 2 different hypotheses: a
cascade hypothesis and a muon hypothesis. Muon and cascade reconstruction
methods are described in
refs.~\cite{ama:mu-reco,kowalski:workshop,ama:b10casc-pub}. We obtain a cascade
vertex resolution of about 6~m in the x,y coordinates and slightly better in
the z coordinate. We obtain a cascade energy resolution of $\log_{10}E_{\rm
  true}/E_{\rm reco} \approx$~0.15. The Rolling analysis reconstructs the 
position of cascades while for the Temporal analysis both the position and the
energy of the cascade is reconstructed. The angular resolution of the muon
reconstruction is about $5^\circ$ \footnote{Better angular resolution is
achieved by analyses that focus on the muon channel.}.

\section{Rolling Time Window Analysis}

The Rolling analysis currently uses data from the year 2001.  We scan the
entire data sample for a clustering of events which survive cuts and are not
consistent with the expected background.  Therefore, it has the potential to
detect signals which are not coincident with prior gamma-ray detections.
These sources include gamma-ray dark neutrino sources, such as choked
GRBs~\cite{choke-bursts} as well as conventional GRBs not detected by the 
Third Interplanetary Network (IPN3)~\cite{ipn3}. 
The live-time of this analysis is $\approx$233 days.  
Two separate rolling searches are performed, with time window lengths of 1
and 100 seconds.  These lengths were chosen, based on the bimodal plot of GRB
durations produced by BATSE~\cite{bimodal}, to contain the majority of signal
from short and long bursts, respectively, while still being short enough to
keep out extra background events.
Since there is no temporal or spatial coincidence to aid in background
rejection, the use of cuts to reduce the background of atmospheric muons
becomes very important.  Cuts based on both topology and number of hits in
optical modules (which is indirectly tied to event energy) are utilized. 
After an initial filter is applied to take only high energy events, a final
cut is made using a support vector machine (SVM)~\cite{svm}. 
Background muon Monte Carlo was found to be in good agreement with
experimental data in all 6 variables used in the SVM. A sample of experimental
data taken from 5 runs distributed throughout the year was used as background
in the SVM. The SVM cut was optimized independently for the 1 second and 100
second searches.
Since the background is of stochastic nature, Poisson statistics can be
used to estimate the the statistical significance of a cluster of
events. Preliminary calculations result in a sensitivity of
$2.7\times$10$^{-6}$~GeV~cm$^{-2}$~s$^{-1}$~sr$^{-1}$ for a time-averaged
diffuse neutrino flux of all flavors, with energy spectrum according to the
Waxman-Bahcall model. This sensitivity assumes 425 bursts during the live-time
of this analysis based on the average rate of GRB detection by the BATSE
experiment and does not account for the unknown number of bursts with weaker
or non-existent gamma-ray signals.  Final results are not yet available at the
time of writing. 

\section{Temporal Analysis: Bursts reported by BATSE in 2000}

\begin{figure}[ht]
\begin{center}
\mbox{
\includegraphics*[width=0.36\textwidth]{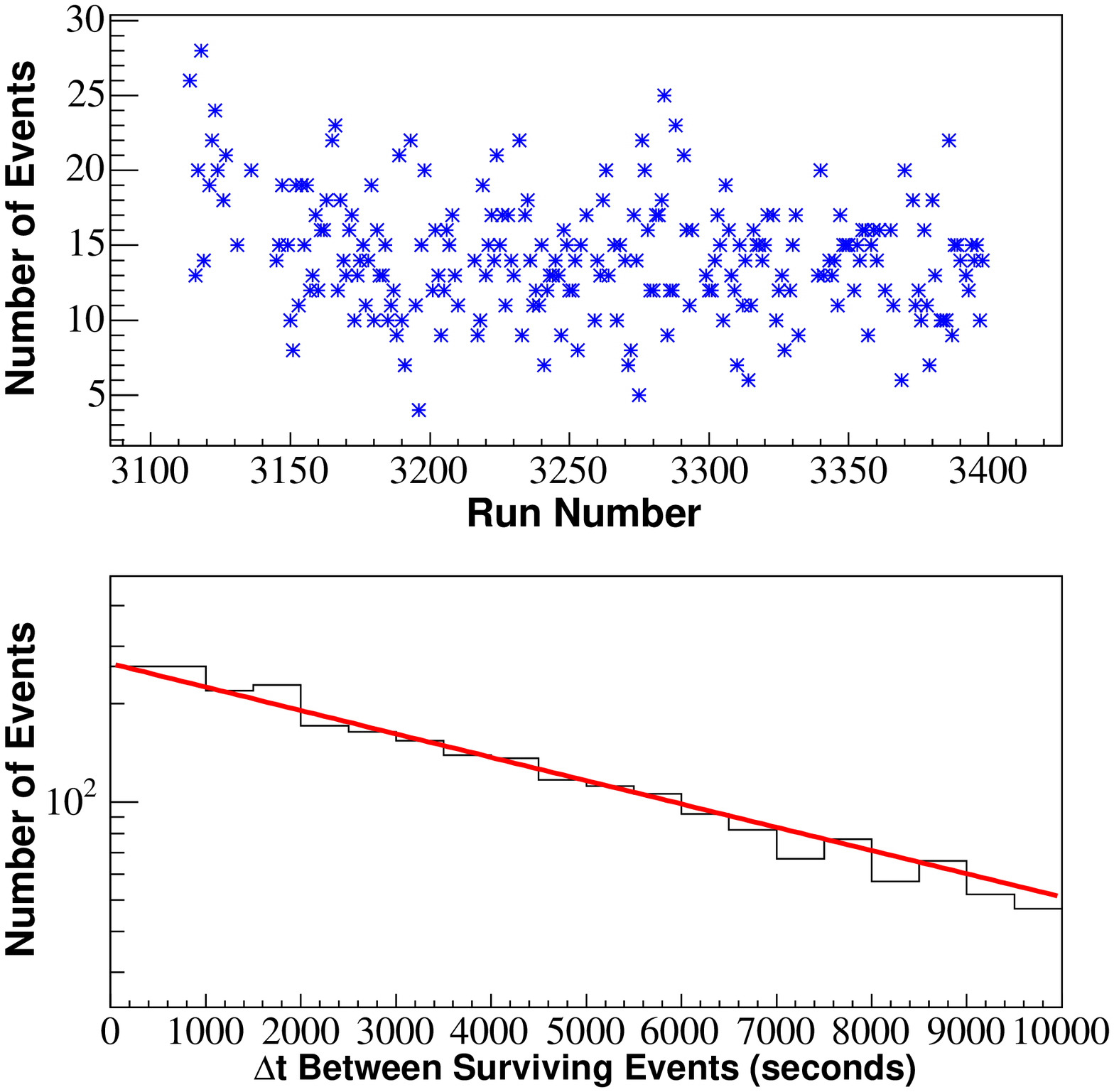}
\includegraphics*[width=0.37\textwidth]{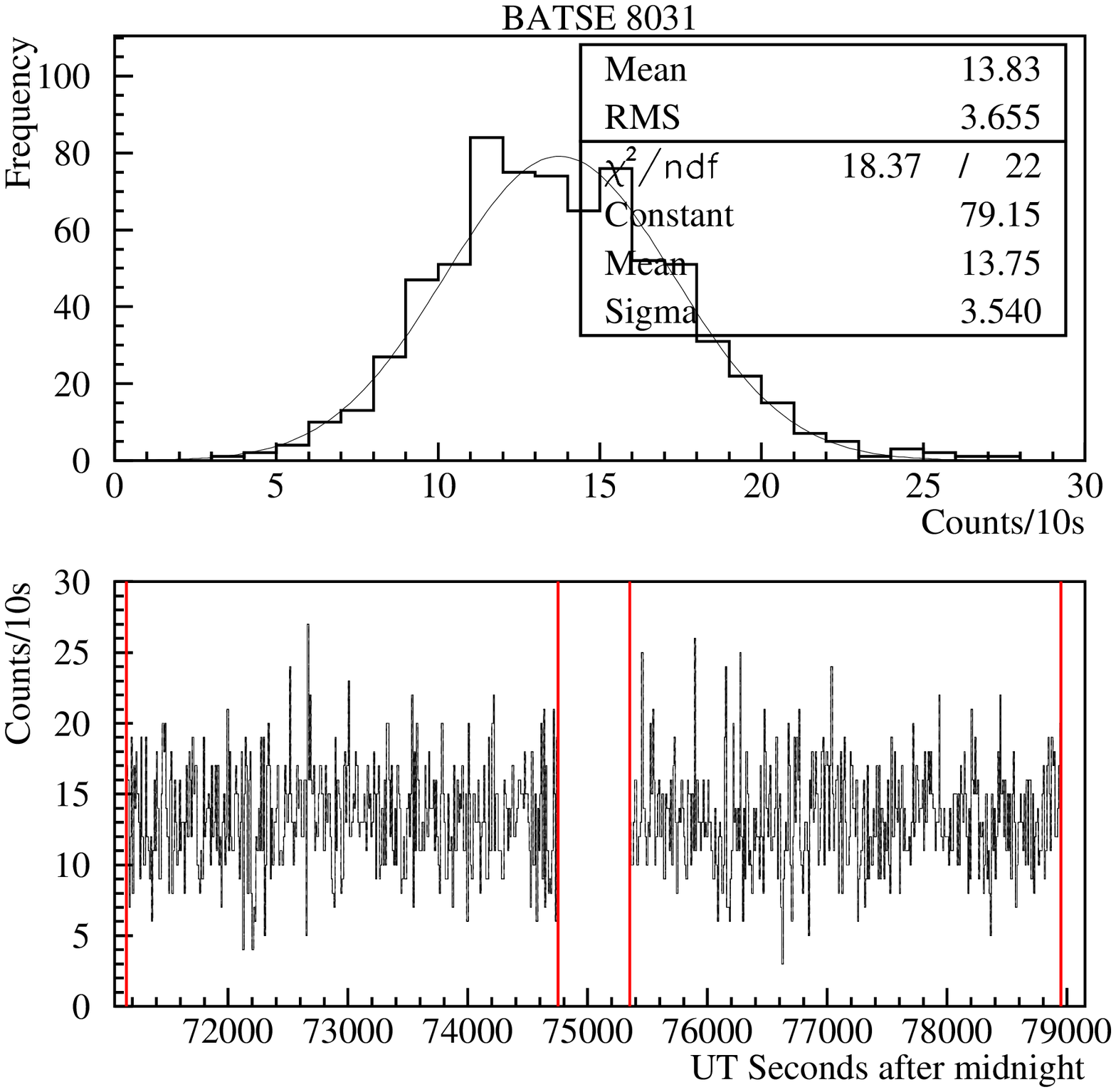}}
\caption{\label{fig:stability} Left - Rolling Analysis:
Plots showing background stability. The upper plot shows background counts per
day after the SVN cut and the lower plot shows time between
events surviving cuts.  The line 
is the prediction assuming Poissonian statistics. Right - Temporal
Analysis: Stability plots for BATSE-8031. The upper panel shows
the frequency of events/10s that pass the filter in the off-time window. The
lower panel shows the number of events/10s that pass the filter versus time in
seconds after midnight (UTC). The vertical lines indicate the off-time
period. The on-time period is analyzed according to our blindness
procedures.}
\end{center}
\end{figure}

AMANDA-II began normal operation Feb. 13, 2000. The last BATSE burst was
reported May 26, 2000. In this period 76 bursts were reported. Since the GRB
start time and duration\footnote{We use $T90$, the time over which a burst 
emits from 5\% to 95\% measured fluence, as the duration}, are well known, the
separation of $\nu$-induced cascade signal from the down-going muon
background is simplified. We use three selection criteria based
on the two reconstruction hypotheses to discard the down-going muon background
and keep the neutrino-induced cascade signal.
A total of $\approx$7800~s per burst were studied. A period of 600 s
(\textit{on-time} window) centered  at the start time of the GRB was initially
set aside in accordance with our blind analysis procedures. Two periods of
data of 1 hour duration (\textit{off-time} window) just 
before and after the on-time window are also studied. We optimize the
selection criteria using the off-time window and signal simulation. Thus the
background is experimentally measured. We only examined the
fraction of the on-time window corresponding to the duration of each
burst. Keeping the rest of the on-time window blind 
allows for other future searches, e.g. precursor neutrinos.
We determined the detector stability using the off-time window
experimental data. Only GRBs for which the detector is found to be stable in
the off-time windows were used. Of the 76 bursts reported
by BATSE in coincidence with AMANDA-II, for two bursts there are gaps in the
AMANDA-II data and for one burst, AMANDA-II data was found to be unstable. Figure
\ref{fig:stability} shows a sample of the plots used to determine the
stability.
We applied the selection criteria in two steps, a filter and final
selection. The filter rejects down-going muons, $\theta_\mu > 
70^\circ$, and keeps events that are cascade-like, $L_{mpe} < 7.8$. The
parameter $L_{mpe}$ is the reduced likelihood of the cascade vertex
reconstruction and has smaller values for cascade-like events.
The final selection criteria are $L_{mpe}<6.9$ and $E>40$~TeV, where $E$ is
the reconstructed cascade energy. One event in the off-time window remains
after all cuts. This is equivalent to a background of
$n_b=0.0054^{+0.013}_{-0.005}$~(stat) in the on-time window.
After un-blinding the on-time window, no events were found. To set a
limit, we assume a WB-like spectrum with $E_b=100$~TeV and $E_s=10$~PeV. We
assume neutrino flavor flux ratio of 1:1:1 and p-$\gamma$ neutrino
generation. The 90\% c.l. limit on the all-flavor flux factor is  
$9.5\times$10$^{-7}$~GeV~cm$^{-2}$~s$^{-1}$~sr$^{-1}$. The event upper limit
is 2.44. These limits have not yet been corrected for systematic
uncertainties. Once the systematic uncertainties have been taken into account
this limit will worsen slightly.

\section{Outlook and Conclusions}

Two methods for searching for neutrino-induced cascades from GRBs using 
AMANDA-II have been presented.  A Rolling Time Window search is being
conducted to search for a neutrino GRB signal at any time and from any
direction.  This method serves as a useful complement to satellite-coincident
GRB searches conducted with AMANDA-II. Its sensitivity to individual bursts
suffers from the lack of temporal constraints, but it has the potential to
observe neutrino signals from transients which would otherwise be missed.
Although this search is currently being conducted on the 2001 data set, it is
relatively straightforward to expand the search to data sets from later years.
This method can also be adapted to use the muon channel in addition to
cascades.

Temporal correlation with satellites was used to perform a search with very
low background. No evidence for neutrinos was found and we have set a limit
based on the WB flux. The 90\% c.l. limit on the all-flavor flux factor,
supposing 1:1:1 flavor flux ratio, is
$9.5\times$10$^{-7}$~GeV~cm$^{-2}$~s$^{-1}$~sr$^{-1}$. This value has not
yet been corrected for systematic uncertainties. Previous
searches by AMANDA-II, performed on a much larger set of
bursts~\cite{icrc03:grb}, have a significantly better sensitivity  
than what has been presented here. Given a large random set of bursts with
both positive and negative declination, we expect the cascade channel
sensitivity to be roughly half as sensitive as the muon channel. It is
expected that only a small fraction of all bursts will contribute
significantly an eventual observed neutrino flux. By monitoring both
hemispheres we increase the probability of discovery. The Temporal Analysis
can be expanded to include bursts reported by IPN3, Swift and by using newly
or soon to be deployed IceCube strings.

\end{document}